\pdfoutput=1
\documentclass[fleqn,usenatbib,usedcolumn]{mnras}

\usepackage{graphicx,amssymb,hyperref}
\usepackage{multirow}
\usepackage{color}

\usepackage[fleqn]{amsmath}
\usepackage{abbrev}
\usepackage{siunitx}
\usepackage{comment} 
\usepackage{gensymb}
\usepackage{booktabs,tabularx}
\usepackage{flushend}
\usepackage{xspace}
\usepackage{expl3}
\usepackage{physics}
 
\usepackage{nicefrac}
\usepackage{units}
\usepackage{verbatim}

\def\mnras{MNRAS}
\def\apj{ApJ}

\def\simlt{\lower.5ex\hbox{$\; \buildrel < \over \sim \;$}}
\def\simgt{\lower.5ex\hbox{$\; \buildrel > \over \sim \;$}}
\def\etal{{\it et al.}}

\def\kpc{\mathrm{\, kpc}}

\def\mpc{\mathrm{\, Mpc}}
\def\gpc{\mathrm{\, Gpc}}
\def\msun{\mathrm{\, M_\odot}}
\def\kms{\mathrm{\, km \, s^{-1}}}
\def\cmsg{\, \mathrm{cm^2 \, g^{-1}}}
\def\gyr{ \, \mathrm{Gyr}}

\newcommand{\eagle}{\textsc{eagle}\xspace}
\newcommand{\ceagle}{\textsc{c-eagle}\xspace}

\def\gs{\mathrel{\raise1.16pt\hbox{$>$}\kern-7.0pt \lower3.06pt\hbox{{$\scriptstyle \sim$}}}}         
\def\ls{\mathrel{\raise1.16pt\hbox{$<$}\kern-7.0pt \lower3.06pt\hbox{{$\scriptstyle \sim$}}}}

\newcommand{\be}{\begin{equation}}
\newcommand{\ee}{\end{equation}}
\newcommand{\ba}{\begin{eqnarray}}
\newcommand{\ea}{\end{eqnarray}}

\DeclareGraphicsExtensions{.pdf,.png}

\title[Galaxy cluster density profiles with SIDM+baryons]{The diverse density profiles of galaxy clusters with self-interacting dark matter plus baryons}

\author[A.\ Robertson \etal]
{\parbox{\textwidth}{Andrew Robertson\thanks{e-mail: {\tt andrew.robertson@durham.ac.uk}},$^1$ 
Richard Massey,$^1$ Vincent Eke,$^1$ Sean Tulin,$^2$  Hai-Bo Yu,$^3$ Yannick Bah\'e,$^{4,5}$ David J.\ Barnes,$^{6,7}$ Richard G.\ Bower,$^1$ Robert A.\ Crain,$^8$ Claudio Dalla Vecchia,$^{9,10}$ Scott T.\ Kay,$^6$ Matthieu Schaller$^{1,4}$ and Joop Schaye$^4$}
\vspace{0.3cm}
\\$^1$Institute for Computational Cosmology, Durham University, South Road, Durham DH1 3LE, UK
\\$^2$Department of Physics and Astronomy, York University, Toronto, Ontario M3J 1P3, Canada
\\$^3$Department of Physics and Astronomy, University of California, Riverside, California 92521, USA
\\$^4$Leiden Observatory, Leiden University, PO Box 9513, NL-2300 RA Leiden, the Netherlands
\\$^5$Max-Planck-Institut f\"ur Astrophysik, Karl-Schwarzschild Str. 1, 85748 Garching, Germany
\\$^6$Jodrell Bank Centre for Astrophysics, School of Physics and Astronomy, The University of Manchester, Manchester M13 9PL, UK
\\$^7$Department of Physics, Kavli Institute for Astrophysics and Space Research, Massachusetts Institute of Technology, Cambridge, MA 02139, USA
\\$^8$Astrophysics Research Institute, Liverpool John Moores University, 146 Brownlow Hill, Liverpool L3 5RF, United Kingdom
\\$^9$Instituto de Astrof\'isica de Canarias, C/ V\'ia L\'actea s/n, 38205 La Laguna, Tenerife, Spain
\\$^{10}$Departamento de Astrof\'isica, Universidad de La Laguna, Av. del Astrof\'isico Francisco S\'anchez s/n, 38206 La Laguna, Tenerife, Spain
}

\begin{document}

\maketitle

\label{firstpage}

\begin{abstract}
We present the first simulated galaxy clusters ($M_{200}>10^{14}\msun$) with both self-interacting dark matter (SIDM) and baryonic physics. They exhibit a greater diversity in both dark matter and stellar density profiles than their counterparts in simulations with collisionless dark matter
(CDM), which is generated by the complex interplay between dark matter self-interactions and baryonic physics. Despite variations in formation history, we demonstrate that analytical Jeans modelling predicts the SIDM density profiles remarkably well, and the diverse properties of the haloes can be understood in terms of their different final baryon distributions.
\end{abstract}


\section{Introduction}

The possibility that dark matter (DM) particles can interact with one another through forces other than just gravity has received significant attention since it was first proposed by \citet{2000PhRvL..84.3760S}. The existence of self-interacting dark matter (SIDM) would have significant implications for both particle physics and astrophysics. A detection of significant self-interactions would rule out many popular DM candidates such as axions \citep{2009NJPh...11j5008D} or supersymmetric neutralinos \citep{2005PhR...405..279B}, and would alter cosmological structure formation on small scales. SIDM would explain why many dwarf and low surface brightness galaxies appear to have less DM in their centres than is predicted when DM particles are assumed to be collisionless \citep{2013MNRAS.430...81R,2014MNRAS.444.3684V,2015MNRAS.453...29E}. It also provides a mechanism to produce a {\it diversity} of DM density profiles \citep{2016arXiv160908626E,2017MNRAS.468.2283C,2017PhRvL.119k1102K} that is difficult to achieve within the standard paradigm \citep{2015MNRAS.452.3650O}, and appears to be necessary if DM density profiles are being correctly inferred from observations \citep[e.g.][]{2008AJ....136.2648D,2008ApJ...676..920K,2014ApJ...789...63A,2015AJ....149..180O}. 

The mechanism by which SIDM alters the density profile of a halo is thermalization. Self-interactions redistribute energy between particles, heating up the centre of the halo, which would otherwise have a low velocity dispersion. These heated particles move to orbits with larger apocenters, shifting mass from the centre to larger radii. Regions where SIDM particles have scattered multiple times approach thermal equilibrium, which has led to the modelling of SIDM as an isothermal gas \citep{2014PhRvL.113b1302K}, known as \emph{Jeans modelling}. A key prediction of this method is that baryons play an important role in determining the final SIDM profile, with slight differences in baryonic distributions leading to very different rotation curves \citep{2017PhRvL.119k1102K}.

The high densities and velocity dispersions in galaxy clusters mean that for a given SIDM cross-section, the thermalisation timescales are shorter than in individual galaxies. These systems can therefore provide strong constraints on the self-interaction properties of DM. At the same time, physical processes within a cluster span a large dynamical range, making them computationally expensive to study using $N$-body simulations. Analytical methods can be applied to these massive systems, but it is important to verify that these methods work.

This letter introduces the first simulations of galaxy clusters to incorporate both SIDM and the baryonic processes that are important for galaxy formation. These simulated clusters provide an ideal way to test constraints placed on the SIDM cross-section from observed clusters, and are used here to explicitly test Jeans modelling of SIDM. We introduce the simulations in \S\ref{sect:Simulations}, before discussing the density profiles of our simulated clusters in \S\ref{sect:density_profiles}. We conclude in \S\ref{sect:conclusions}.

\section{Simulations}
\label{sect:Simulations}

\subsection{The Cluster-EAGLE simulations}
\label{sect:C-EAGLE}

We have re-simulated two clusters from the Cluster-EAGLE (\ceagle) project \citep{2017MNRAS.470.4186B,2017MNRAS.471.1088B}: CE-05 and CE-12, with masses of $M_{200} = 1.4$ and $\num{3.9e14} \msun$ respectively.\footnote{We define $r_{200}$ as the radius at which the mean enclosed density is 200 times the critical density, and $M_{200}$ as the mass within $r_{200}$.}  
Both clusters are classified as `relaxed', based on their gas properties at $z=0.1$ \citep{2017MNRAS.471.1088B}.
We ran four simulations of each cluster:
CDM-only, CDM+baryonic physics, SIDM-only and SIDM+baryonic physics. 
An isotropic and velocity-independent cross-section of $\sigma / m=1 \cmsg$ was used for all runs with SIDM. 

The \ceagle project uses the zoom simulation technique \citep{1993ApJ...412..455K} to resimulate (at higher resolution) galaxy cluster haloes found in a parent simulation with a side length of $3.2 \gpc$ \citep{2017MNRAS.465..213B}. The high-resolution region around each cluster is selected so that no lower resolution particles are present within a radius of $5 \, r_{200}$ from the cluster centre at $z = 0$. The high-resolution region matches the resolution of the \eagle$100 \mpc$ simulation \citep[Ref-L100N1504,][]{2015MNRAS.446..521S}, with DM particle mass $m_\mathrm{DM} = \num{9.7e6} \msun$ and initial gas particle mass $m_\mathrm{gas} = \num{1.8e6} \msun$. Runs including baryons used the \eagle galaxy formation model \citep{2015MNRAS.446..521S,2015MNRAS.450.1937C}, which includes radiative cooling, star formation, stellar evolution, feedback due to stellar winds and supernovae, and the seeding, growth and feedback from black holes. 
The specific calibration of \eagle that was used, is labelled as `\emph{AGNdT9}' in \citet{2015MNRAS.446..521S}. This was chosen as it provides a better match to the observed gas fraction and X-ray luminosity--temperature relation of galaxy groups than the fiducial `\emph{Ref}' calibration. All the simulations used a \emph{Planck} 2013 cosmology \citep{2014A&A...571A..16P}.\footnote{Specifically, $\Omega_\mathrm{b} = 0.04825$,  $\Omega_\mathrm{m} = 0.307$, $\Omega_{\Lambda} = 0.693$, $H_0 = 67.77 \kms \mpc^{-1}$, $\sigma_8 = 0.8288$, $n_\mathrm{s} = 0.9611$ and $Y = 0.248$.}

Because this work focusses on radial density profiles, we here summarise the strengths and weaknesses of \ceagle in this respect. 
\ceagle clusters simulated in a CDM universe have total stellar content and black hole masses that match observed relations
\citep{2017MNRAS.470.4186B}, but their central galaxies are $\approx 3$ times too massive.
The simulated clusters are slightly too gas rich overall, but have a deficit of gas in their centres, where the gas is also too hot \citep{2017MNRAS.471.1088B}. 
Star particles in \eagle lose mass to the surrounding gas, which can lead to the formation of massive gas particles in gas-poor regions. We have found that both the CDM and SIDM versions of CE-05 form a single massive gas particle at the centre of the halo. The density profile of this particle, using its SPH kernel, is shown as a dashed line 
in Figure~\ref{fig:density_profiles}.

\begin{table*}
\centering
\begin{tabular}{llccccccccccc}
\hline 
\\[-9pt]
Halo  & Physics  & $M_{200}$ & $r_{200} $ & $M_{*}$ & $M_\mathrm{gas}$ & $f_\mathrm{bar}$ & \!\!$c_{200}$\!\! & $r_\mathrm{b} $ & $\lambda'$ & \!\!\!$M_{*}(<$$30 \kpc)$\!\! & \!\!$M_\mathrm{DM}(<$$30 \kpc)$\!\! & \!\!$t^*_{1/2}$\!\!  \\[2pt]
  &     & \!\![$10^{14}\msun$]\!\! & \!\!\![$\mpc$]\!\!\! & \!\![$10^{12} \msun$]\!\! & \!\![$10^{12}\msun$]\!\! & [$\%$] &  & \!\![$\kpc$]\!\! &  & \!\![$10^{12} \msun$] & \!\![$10^{12} \msun$]\!\! & \!\![$\gyr$]\!\!  \\[2pt] \hline
CE-05\!\!\!\!\!\! & CDM          & 1.37   & 1.09             &                 &                          &          & 6.40  &    & 0.027     \\
      & CDM+baryon\!\!\!\!\!\!  & 1.38   & 1.09             & 1.86 & 13.7          & 10.5          &      &     &  0.027        & 0.49  & 1.48            &   3.8    \\
      & SIDM         & 1.36   & 1.08             &                 &                          &                  &     & 95      & 0.028            \\
      & SIDM+baryon\!\!\!\!\!\! & 1.36  & 1.09             & 1.95 & 12.9          & 10.9         &     &      & 0.029  & 0.61   & 1.02             & 4.3     \\ \hline
CE-12\!\!\!\!\!\! & CDM          & 3.92   & 1.54             &                 &                          &                  & 4.35   &   & 0.036    \\
      & CDM+baryon\!\!\!\!\!\! & 3.96   & 1.55             & 5.74 & 53.8          & 15.0          &     &      & 0.040       & 0.58   & $1.21$            & 6.8 \\
      & SIDM         & 3.88  & 1.54             &                 &                          &                  &      & 199     & 0.036    \\
      & SIDM+baryon\!\!\!\!\!\! & 3.91   & 1.54             & 5.85 & 52.2          & 14.9          &      &     &  0.039        & 0.24   & 0.26        & 5.3  
\end{tabular}
\caption{The $z = 0$ properties of the two halos, in each of the four Physics runs. $M_{*}$ and $M_\mathrm{gas}$ are the total stellar and gas masses within $r_{200}$, while $f_\mathrm{bar}$ is the baryon fraction within $r_{200}$. The concentration, $c_{200}$, of each halo was only calculated for the CDM-only simulations. Similarly, the Burkert scale radius, $r_\mathrm{b} $, was only calculated for SIDM-only simulations. $\lambda'$ is the halo spin parameter, including all mass within $r_{200}$. $M_{*}(<$$30 \kpc)$ and $M_\mathrm{DM}(<$$30 \kpc)$\ are measurements of the stellar and DM mass within a $30 \kpc$ spherical aperture, centred on the most bound particle. $t^*_{1/2}$ is the age of the universe when $M_{*}(<$$30 \kpc)$ was 50\% of its $z = 0$ value. }
\label{simulation_table}
\end{table*}

\subsection{Implementation of dark matter scattering}

During each simulation time-step, DM particles 
search for neighbours within a radius $h_\mathrm{SI}$, and scatter 
isotropically with probability
\begin{equation}
\label{eq:P_ij}
P_\mathrm{scat}= \frac{(\sigma/m) ~ m_\mathrm{DM} \, v \, \Delta t}{\frac{4\pi}{3} h_\mathrm{SI}^{3}},
\end{equation}
where 
$v$ is the particles' relative velocity, and $\Delta t$ is the size of the time-step \citep{2017MNRAS.465..569R}.
Provided it is smaller than resolved structures, the results are insensitive to the exact choice of $h_\mathrm{SI}$ \citep{myphd}. 
We therefore fix $h_\mathrm{SI}$ to a constant comoving size of $2.66 \kpc$, matching the gravitational softening length in \eagle before $z=2.8$.

\subsection{Structural properties of simulated clusters}
\label{sect:sim_suite}

The properties of the clusters at $z$$=$$0$ are listed in Table~\ref{simulation_table}. As well as the total, stellar and gas mass within $r_{200}$, we show the parameters of a fit to the density profiles of our DM-only runs. For CDM, we fit a \citet[][NFW]{1997ApJ...490..493N} profile \!\!
\begin{equation}
\label{eq:rho_NFW}
\frac{\rho_\mathrm{NFW}(r)}{\rho_\mathrm{crit}} = \frac{\delta_\mathrm{NFW}}{(r/r_\mathrm{s})(1+r/r_\mathrm{s})^2},
\end{equation}
where $r_\mathrm{s}$ is a scale radius, $\delta_\mathrm{NFW}$ a dimensionless characteristic density, and $\rho_\mathrm{crit}=3H^2/8\pi G$ is the critical density. The NFW concentration parameter is defined as $c_{200} \equiv r_{200}/r_\mathrm{s}$. 
For SIDM runs, we fit a \citet{1995ApJ...447L..25B} profile
\begin{equation}
\label{eq:rho_burkert}
\rho_\mathrm{B}(r) = \frac{\rho_\mathrm{b} \, r_\mathrm{b}^3}{(r + r_\mathrm{b}) (r^2 + r_\mathrm{b}^2)},
\end{equation}
which has a constant density core inside radius $r_\mathrm{b}$.
All fits 
were performed 
between radii 
$0.01 \, r_{200}$ and $r_{200}$, minimising 
\begin{equation}
\sum_{i=1}^{N_\mathrm{bins}} \left( \log \rho_\mathrm{sim} (r_i) - \log \rho_\mathrm{fit} (r_i) \right)^2,
\end{equation}
where the $N_\mathrm{bins}=50$ radial bins are logarithmically spaced. 
The CDM-only profiles are well fit at radii outside $10\kpc$, but exceeded the NFW model in the very centre.
Both SIDM-only profiles are well fit on all scales \citep[see also][]{2013MNRAS.430...81R}.

For each simulated halo, we also calculate 
the \cite{2001ApJ...555..240B} dimensionless spin parameter
\begin{equation}
\label{eq:bullock_spin}
\lambda' \equiv \frac{|\mathbfit{J}|}{\sqrt{2} M_{200} V_{200} r_{200}},
\end{equation}
where $\mathbfit{J}$ is the angular momentum of all mass within $r_{200}$, about the most-bound particle. The velocities 
in $\mathbfit{J}$
are with respect to the mass-weighted average velocity of the halo.

\section{Halo density profiles}
\label{sect:density_profiles}

\subsection{Simulation results}

\begin{figure*}
        \centering
        \begin{tabular}{cc}
\includegraphics[width=\columnwidth,trim={0 12mm 0 0},clip]{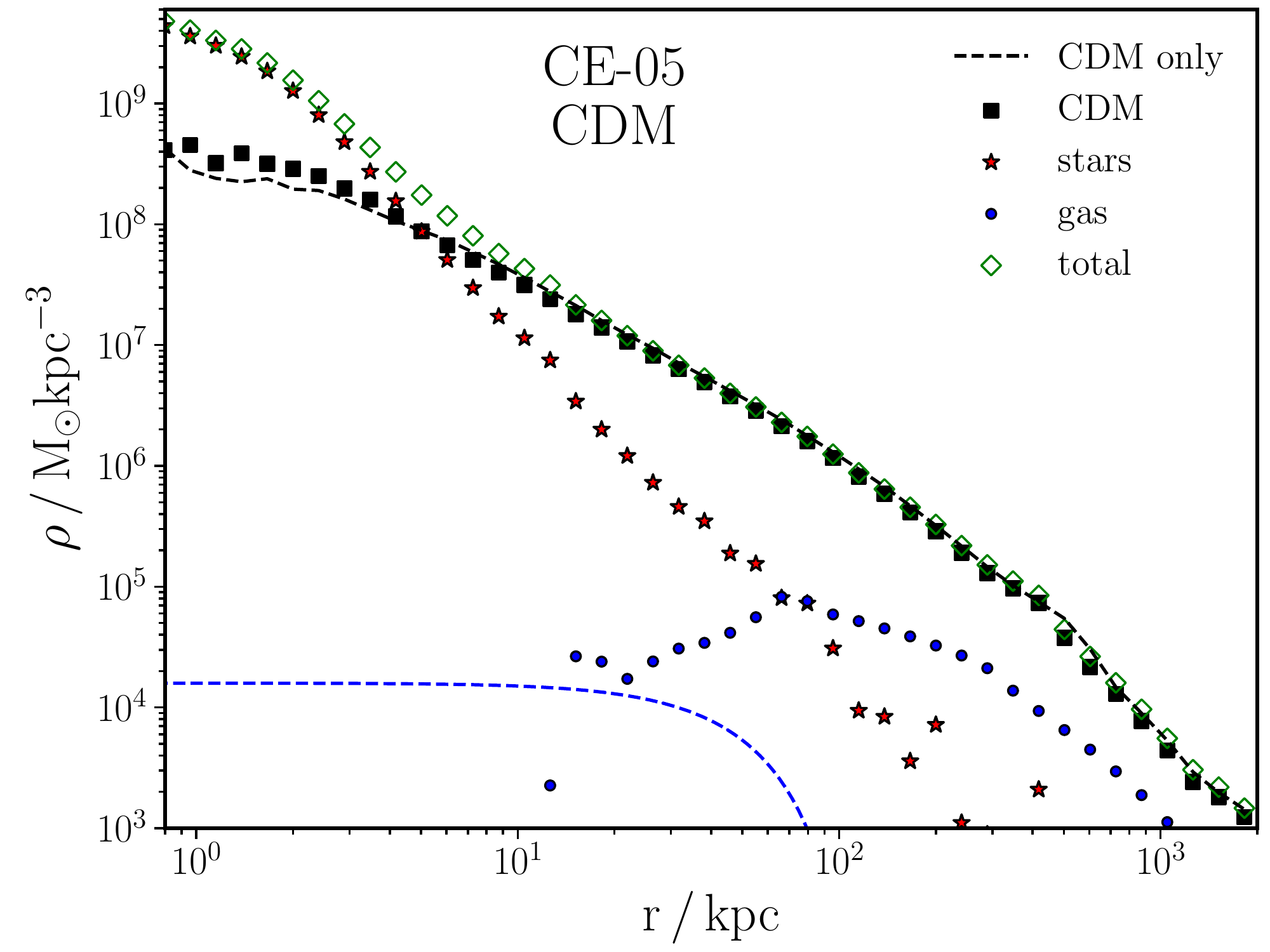} &  ~
\includegraphics[width=\columnwidth,trim={0 12mm 0 0},clip]{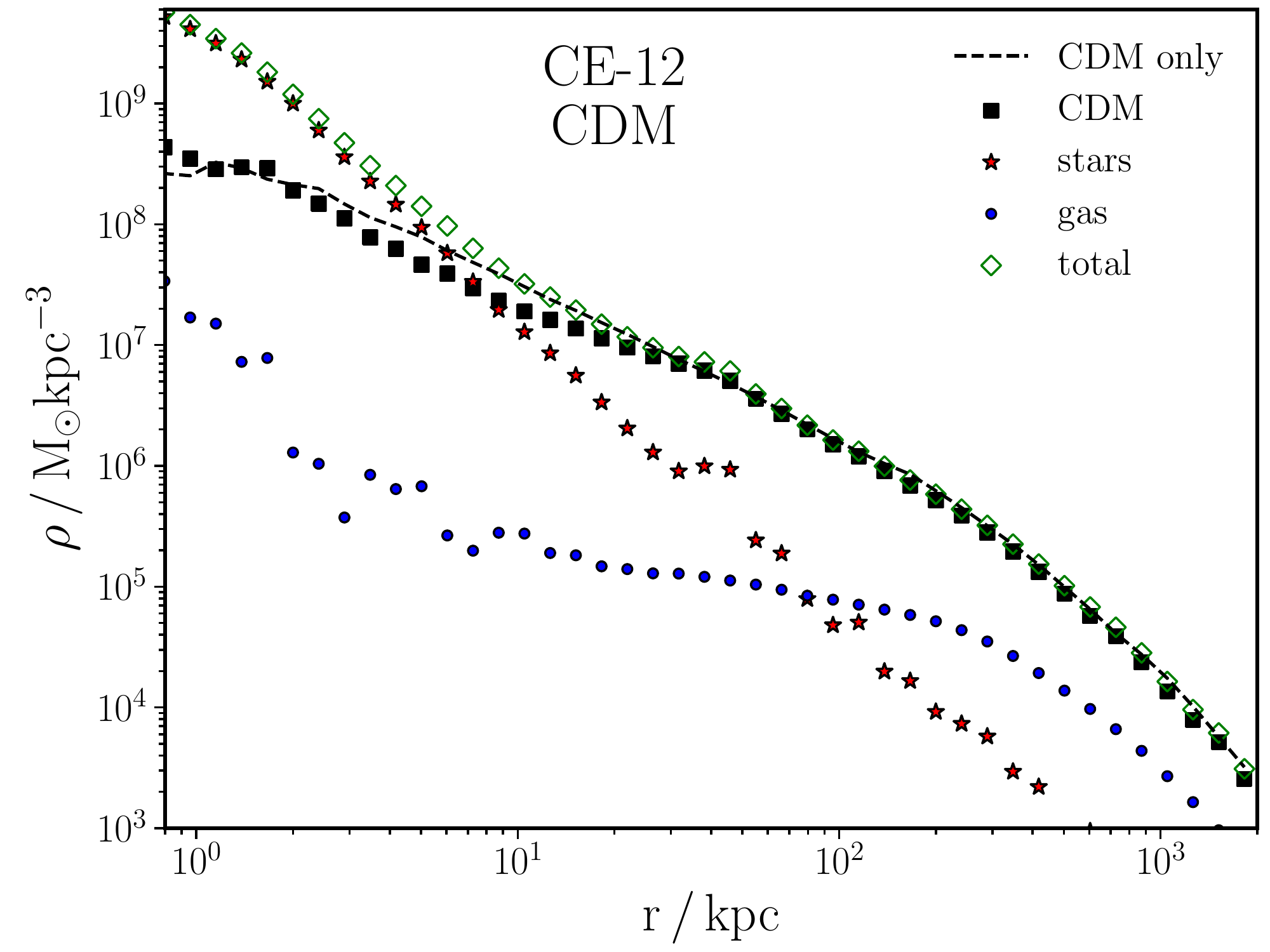} \\ 
\includegraphics[width=\columnwidth,trim={0 1mm 0 1mm},clip]{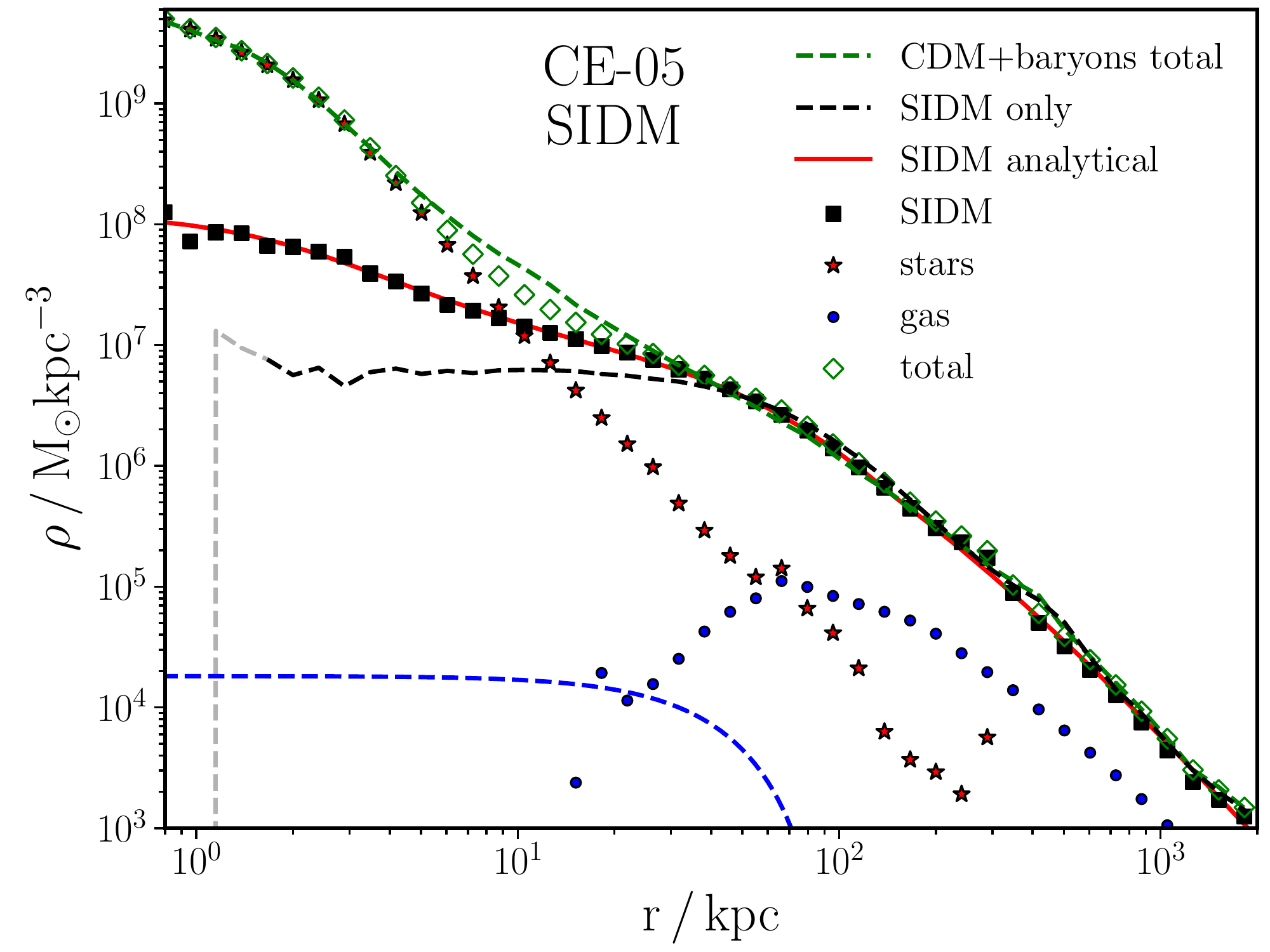} &  ~
\includegraphics[width=\columnwidth,trim={0 1mm 0 1mm},clip]{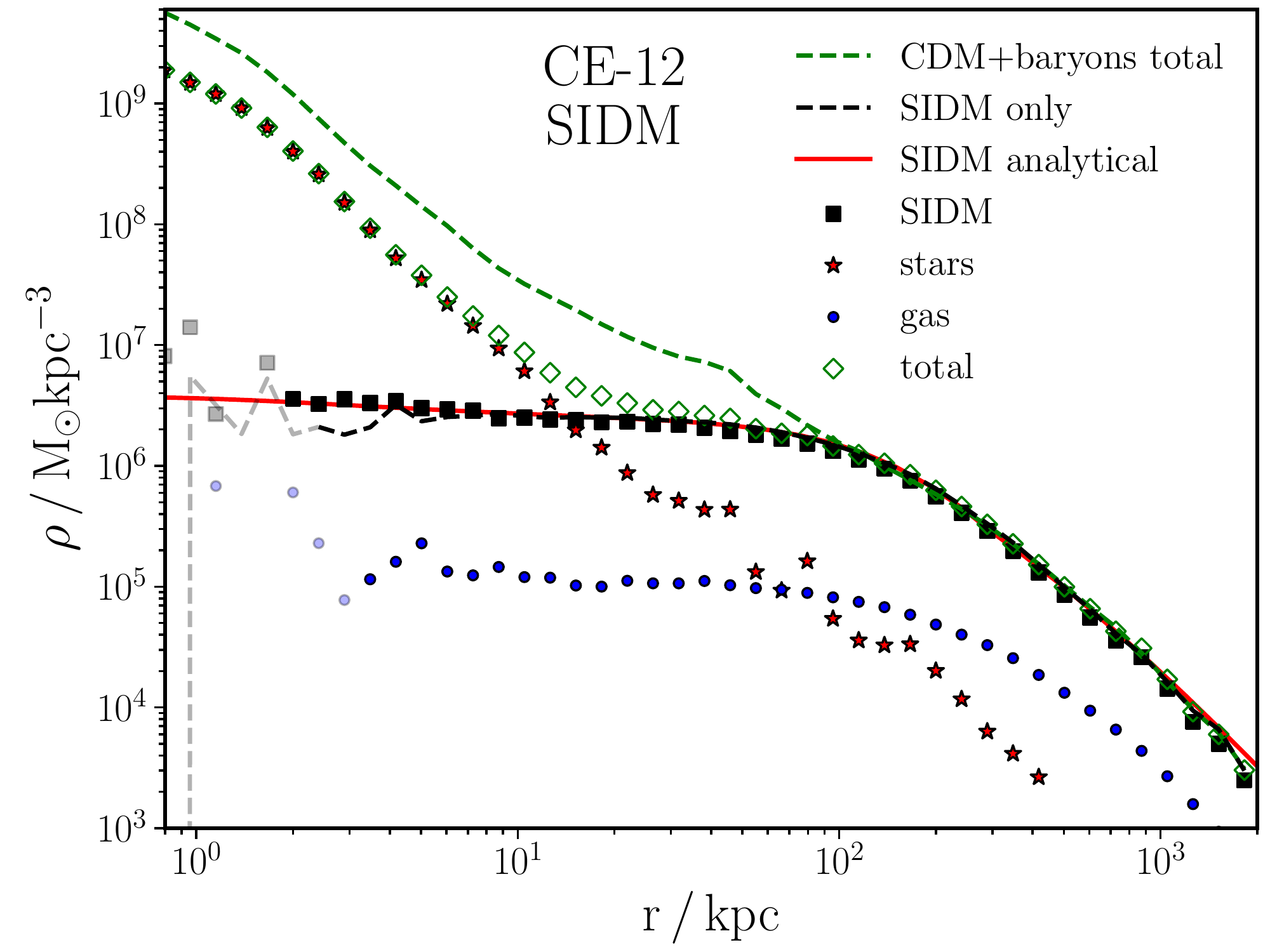} \\
\end{tabular}
	\caption{The $z = 0$ radial density profiles of CE-05 (left) and CE-12 (right), with CDM (top) and SIDM (bottom). The DM-only density profiles are overlaid as black dashed lines, and the SIDM panels have the total density profiles from their CDM counterparts overlaid as green dashed lines. Symbols and lines become semi-transparent when the density corresponds to fewer than 10 particles per radial bin. The solid red lines in the bottom panels are the analytical predictions for the SIDM density, discussed in Section~\ref{sect:analytic}. The blue dashed lines are the density profiles of single massive gas particles that form at the centre of CE-05 (see Section~\ref{sect:C-EAGLE}).}
	\label{fig:density_profiles}
\end{figure*}

When baryons are added to CDM simulations (top panels of Figure~\ref{fig:density_profiles}), stars dominate the central 10$\kpc$ of the total mass profile, but the DM density profile 
stays almost unchanged. 
When baryons are added to SIDM haloes (bottom panels), the 
response of the two clusters is starkly different.
The SIDM halo of CE-12 develops a large core of constant DM density, with or without baryons.
The density of stars in its inner $20 \kpc$ is 
less than half of that with CDM+baryons, but with a similar radial dependence.
On the other hand, including baryons in CE-05 enhances the central DM density relative to the SIDM-only run, removing the constant density core and recovering a cuspy  total density profile that differs only slightly 
from that with CDM+baryons. 

\subsection{Semi-analytic Jeans modelling of SIDM}
\label{sect:analytic}

The contrast between the cored SIDM profile of CE-12, which is unaffected by baryons, versus the creation of an SIDM cusp in the baryonic version of CE-05, is a consequence of the two clusters' different baryonic distributions. \citet{2016PhRvL.116d1302K} successfully fit the rotation curves of simulated SIDM halos using a model where SIDM behaves as an isothermal gas within the radius (known as $r_1$) at which particles have scattered once over the age of the halo. The density profile in this isothermal regime is predicted by solving Poisson's equation, while requiring hydrostatic equilibrium.\footnote{The temperature of this `isothermal gas', is related to the SIDM velocity dispersion, such that the SIDM follows the ideal gas law $p = \rho \, \sigma_0^2$, where $p$ and $\rho$ are the SIDM pressure and density and $\sigma_0$ is the one-dimensional velocity dispersion.} The results of this procedure (using $M_{200}$ and $c$ from the CDM-only simulations, the baryon distribution from SIDM+baryons, and the true cross-section of $\sigma/m = 1 \cmsg$ as inputs) are shown in Figure~\ref{fig:density_profiles}, and are an excellent match to the SIDM density measured in both SIDM+baryons simulations.

In the inner regions of our haloes, where the baryons dominate, this analytical prescription leads to a DM density profile \citep{2014PhRvL.113b1302K}
\begin{equation}
\label{eq:isothermal_dens_profile}
\rho_\mathrm{DM}(r) \approx \rho_0 \exp \left\{ \frac{ \Phi_\mathrm{B}(0) - \Phi_\mathrm{B}(r) }{\sigma_0^2} \right\},
\end{equation}
where $\rho_0$ is the central DM density, $\Phi_\mathrm{B}$ is the gravitational potential due to the baryonic mass distribution, and $\sigma_0$ is the one-dimensional velocity dispersion of SIDM (which is approximately constant inside $r_1$, because DM interactions efficiently redistribute energy between SIDM particles).

From equation~\eqref{eq:isothermal_dens_profile} we can see that the density in the inner regions of the DM halo will be roughly constant if $|\Phi_\mathrm{B}(0)| < \sigma_0^2$, while it will increase towards smaller radii if the baryonic potential is significant compared with the DM velocity dispersion. For halos that have been thermalised by DM self-interactions, the central velocity dispersion is roughly the maximum (across all radii) velocity dispersion that the halo would have in the absence of self-interactions \citep[see Figure~6 of][]{2013MNRAS.430...81R}. This in turn is about $0.66 \, v_\mathrm{max}$, independent of the halo mass or concentration \citep{2001MNRAS.321..155L}, where $v_\mathrm{max} \equiv \max \left\{ \sqrt{G M(<r) / r} \right\}$ is the maximum circular velocity of the halo.
 
For CE-05, $v_\mathrm{max} = 848 \kms$ and $\sqrt{|\Phi_\mathrm{B}(0)|} = 1050 \kms$, while for CE-12, $v_\mathrm{max} = 1107 \kms$ and $\sqrt{|\Phi_\mathrm{B}(0)|} = 800 \kms$.\footnote{$\Phi_\mathrm{B}(r)$ was calculated from the radial density profiles of stars and gas, assuming spherical symmetry.} The different behaviour of SIDM in CE-05 and CE-12 is therefore readily understood as a result of the much deeper baryonic potential well in CE-05, combined with CE-12 being more massive (and so with higher DM velocity dispersion). 
Importantly, Jeans modelling produces an excellent match to the DM density in \emph{both} SIDM+baryons simulations, adding 
credence to cross-sections inferred from observational data using this method \citep{2016PhRvL.116d1302K,2017PhRvL.119k1102K}.

\subsection{Discussion}
\label{sec:discussion}

While the SIDM density profiles can be explained in light of the associated baryonic potentials, it is not clear what gave rise to these two haloes having quite different central stellar distributions. After all, the density profiles of the stars in the CDM+baryons versions of CE-05 and CE-12 are similar to one another and to that in the SIDM+baryons version of CE-05. The interesting question is then why SIDM has a large effect on the stellar distribution only in CE-12.

As can be seen in Table~\ref{simulation_table}, neither halo has particularly unusual structural parameters. The concentrations of the CDM-only CE-05 and CE-12 are 6.4 and 4.4 respectively. Given their masses, this places them $0.7 \, \sigma$ above and $0.4 \, \sigma$ below the \cite{2015MNRAS.452.1217C} concentration-mass relation, assuming concentrations to be log-normally distributed with $\sigma_{\log_{10} c} = 0.1$ \citep{2004A&A...416..853D}. The halo spins are also unremarkable: 
the $\lambda'=0.027$ and $0.036$ values from our CDM-only simulations are 
within the typical scatter seen in larger 
simulations, which have 
median $\lambda' \approx 0.035$ independent of halo mass, and 
$\sigma_{\log_{10} \lambda'} \approx 0.22$ \citep[e.g.][]{2001ApJ...555..240B,2007MNRAS.378...55M}.

Any differences between the $z=0$ properties of haloes with SIDM+baryons must ultimately be traceable back to the initial conditions, and should therefore show up in the CDM-only runs. However, even with CDM, the concentration, spin, sphericity, triaxiality, substructure and environment of a halo cannot fully explain the scatter in stellar masses at fixed halo mass \citep{2017MNRAS.465.2381M}.
One notable difference between the haloes is that CE-05 undergoes a 6:1 mass merger at $z \approx 0.2$. However, the qualitative features of CE-05's density profile are the same at $z=0.3$ (before the merger has taken place) as at $z=0$. 
This merger therefore does not affect our conclusions.

Turning to the CDM+baryons runs, there are 
significant differences in the timescale for the build-up of stellar mass within the central galaxy (the BCG). 
At $z=0$, the clusters have a similar stellar mass within a $30 \kpc$ spherical aperture.
For CE-05, half of that stellar mass was already inside this aperture $3.8 \gyr$ after the Big Bang. The same milestone was reached $3 \gyr$ later in CE-12.
SIDM interactions take time to influence the structure of a halo, and are unable to significantly do so in the presence of a deep baryonic potential well. A BCG that builds up its stellar mass early,  
may therefore 
resist the effect of DM interactions to reduce the central DM density. 
Coupled to the fact that CE-12 is a more massive cluster, with a correspondingly larger SIDM temperature, and so more resilience to the inclusion of baryons, the different early formation histories of these haloes may explain why they react so differently to the inclusion of DM interactions by $z=0$. 

Importantly, the SIDM+baryons version of CE-12 does not contain fewer stars than the CDM+baryons version -- 
just fewer stars in the central galaxy. This could be a result of the reduced dynamical friction in a cored density profile \citep{2006MNRAS.373.1451R,2015MNRAS.454.3778P}, leading to less accretion of stellar mass from satellite galaxies.

\section{Conclusions}
\label{sect:conclusions}

We have produced the first simulations of galaxy clusters to include both SIDM and baryonic physics. Of our two simulated SIDM+baryon clusters, one has a large constant density DM core, while the other has a cuspy DM density profile more reminiscent of CDM. We demonstrated that the analytical model introduced in \citet{2014PhRvL.113b1302K} can successfully explain the behaviour of our two haloes: the cuspy SIDM halo has a deep baryonic potential, while the cored halo belongs to a system with a much shallower baryonic potential. What is responsible for these clusters having different central distributions of baryons (mainly stars) is hard to infer from only two simulated systems, but we speculate that the early formation time of the central galaxy in one of our clusters may explain why it is relatively unaffected by DM self-interactions.

Ultimately, the different response between the two haloes to the inclusion of SIDM may not be simply related to the parameters of the CDM haloes, or to their formation history. The complex interplay at the centre of a cluster between gas cooling, gravitational collapse and AGN feedback is highly non-linear, and the behaviour of SIDM in the presence of a dense or diffuse baryonic component provides a positive feedback mechanism.  
Less dense systems are made even less dense by DM self-interactions, while dense systems are relatively unaffected by these interactions. The large differences between our two systems 
suggest that mapping the full diversity of cluster haloes with SIDM will require simulations of more systems (including baryons).

Encouragingly, analytical Jeans modelling of SIDM 
reproduces the density profiles in the simulations.
Even though the distribution of baryons and SIDM 
are hard to predict from the CDM properties of a halo, 
knowledge of the baryon distribution allows one to predict the SIDM distribution for a given cross-section.
This is relevant for the interpretation of observed systems, where the baryon distribution can be observed, and (for different SIDM cross-sections) the predicted DM distributions can be compared with kinematic and gravitational lensing data to infer the best fitting self-interaction cross-section.

\section*{Acknowledgments}
This work was made possible by Lydia Heck's technical support and expertise. Thanks also to Subir Sarkar, Felix Kahlhoefer, Kai Schmidt-Hoberg and Torsten Bringmann for organising an excellent SIDM workshop in Copenhagen, and Manoj Kaplinghat, David Harvey and Mathilde Jauzac for helpful discussions.

This work was supported by the UK Science and Technology Facilities Council, through grants ST/K501979/1, ST/L00075X/1 and ST/L000768/1. It used the DiRAC Data Centric system at Durham University, operated by the Institute for Computational Cosmology on behalf of the STFC DiRAC HPC Facility (www.dirac.ac.uk). This equipment was funded by BIS National E-infrastructure capital grant ST/K00042X/1, STFC capital grants ST/H008519/1 and ST/K00087X/1, STFC DiRAC Operations grant  ST/K003267/1 and Durham University. 
This project also received funding from the EU 
Horizon 2020 research and innovation programme under Marie Sk{\l}odowska-Curie grant agreement 747645.
The C-EAGLE simulations were in part performed on the German federal maximum performance computer ``Hazel Hen'' at the maximum performance computing centre Stuttgart (HLRS), under project GCS-HYDA / ID 44067 financed through the large-scale project ``Hydrangea'' of the Gauss Centre for Supercomputing. Further simulations were performed at the Max Planck Computing and Data Facility in Garching, Germany.
RM and RC were supported by the Royal Society, 
and HBY by the 
US Department of Energy (grant de-sc0008541) and the Hellman Fellows Fund. CDV acknowledges financial support from the Spanish Ministry of Economy and Competitiveness (MINECO) through grants AYA2014-58308 and RYC-2015-1807.

\bibliographystyle{mnras}

\bibliography{bibliography}

\bsp
\label{lastpage}

\end{document}